\documentclass[11pt,a4paper]{article}
\usepackage{amsmath}
\usepackage{amsfonts,amssymb}
\begin{document}

\newcommand{\nl}{\nonumber\\}
\newcommand{\nnl}{\nl[6mm]}
\newcommand{\nle}{\nl[-2.3mm]\\[-2.3mm]}
\newcommand{\nlb}[1]{\nl[-2.3mm]\label{#1}\\[-2.3mm]}

\renewcommand{\theequation}{\thesection.\arabic{equation}}
\let\ssection=\section
\renewcommand{\section}{\setcounter{equation}{0}\ssection}

\newcommand{\be}{\bes}
\newcommand{\ee}{\ees}
\newcommand{\bes}{\begin{eqnarray}}
\newcommand{\ees}{\end{eqnarray}}
\newcommand{\eens}{\nonumber\end{eqnarray}}

\renewcommand{\/}{\over}
\renewcommand{\d}{\partial}
\renewcommand{\div}{\hbox{div}}

\newcommand{\wwedge}{\!\wedge\!}
\newcommand{\ootimes}{\!\otimes\!}
\newcommand{\ooplus}{\!\oplus\!}
\newcommand{\stimes}{\boxtimes}

\newcommand{\zero}{{\barrr 0}}
\newcommand{\one}{{\barrr 1}}

\newcommand{\eps}{\epsilon}
\newcommand{\vth}{\vartheta}
\renewcommand{\th}{\theta}
\newcommand{\dlt}{\delta}
\newcommand{\ups}{\upsilon}
\newcommand{\al}{\alpha}
\newcommand{\bt}{\beta}
\newcommand{\gm}{\gamma}
\newcommand{\om}{\omega}
\newcommand{\si}{\sigma}
\newcommand{\la}{\lambda}
\newcommand{\rep}{\varrho}

\newcommand{\du}{du}
\newcommand{\dz}{dz}
\newcommand{\dt}{dt}
\newcommand{\dth}{d\th}
\newcommand{\dvth}{d\vth}

\newcommand{\dmu}{\d_\mu}
\newcommand{\dnu}{\d_\nu}

\newcommand{\yj}{\eta^j}
\newcommand{\yk}{\eta^k}

\newcommand{\tD}{\widetilde D}
\newcommand{\tQ}{\widetilde Q}
\newcommand{\tR}{\widetilde R}
\newcommand{\tE}{\widetilde E}
\newcommand{\toj}{\tilde\oj}
\newcommand{\boj}{\bar\oj}

\newcommand{\LL}{{L}}
\renewcommand{\L}{{\cal L}}
\newcommand{\J}{{\cal J}}
\renewcommand{\S}{{\cal S}}
\newcommand{\G}{{\cal G}}

\newcommand{\oj}{{\mathfrak g}}
\newcommand{\tr}{\hbox{tr}\,}
\newcommand{\half}{\hbox{$1\/2$}}
\newcommand{\mhalf}{\hbox{$1\/2$}(-)^X}
\newcommand{\third}{\hbox{$1\/3$}}
\newcommand{\mthird}{\hbox{$1\/3$}(-)^X}
\newcommand{\fourth}{\hbox{$1\/4$}}
\newcommand{\mfourth}{\hbox{$1\/4$}(-)^X}
\newcommand{\fifth}{\hbox{$1\/5$}}
\newcommand{\sixth}{\hbox{$1\/6$}}
\newcommand{\msixth}{\hbox{$1\/6$}(-)^X}
\newcommand{\eighth}{\hbox{$1\/8$}}

\newcommand{\ssg}{sl(3)\ooplus sl(2)\ooplus gl(1)}
\newcommand{\vect}{{\mathfrak{vect}}}
\newcommand{\svect}{{\mathfrak{svect}}}
\newcommand{\map}{{\mathfrak{map}}}
\newcommand{\mb}{{\mathfrak{mb}}}
\newcommand{\ksle}{{\mathfrak{ksle}}}
\newcommand{\bigksle}{\overline{{\mathfrak{ksle}}}(5|10)}
\newcommand{\vle}{{\mathfrak{vle}}}
\newcommand{\kas}{{\mathfrak{kas}}}
\newcommand{\vas}{{\mathfrak{vas}}}
\newcommand{\as}{{\mathfrak{as}}}
\newcommand{\kk}{{\mathfrak{k}}}
\newcommand{\fle}{{\mathfrak {le}}}
\newcommand{\sle}{{\mathfrak {sle}}}
\newcommand{\ko}{{\mathfrak {ko}}}
\newcommand{\sko}{{\mathfrak {sko}}}

\newcommand{\brep}[1]{{\bf{\underline{#1}}}}

\newcommand{\barr}{\be\begin{array}}
\newcommand{\earr}{\end{array}}

\newcommand{\lbra}{{\Bbb[}}
\newcommand{\rbra}{{\Bbb]}}
\newcommand{\lcur}{{\Bbb\{}}
\newcommand{\rcur}{{\Bbb\}}}

\newcommand{\into}{\hookrightarrow}
\newcommand{\onto}{\rightarrow}
\newcommand{\hbreak}{\hfill\break}
\newcommand{\larroww}[1]{{\ \stackrel{#1}{\longleftarrow}\ }}

\newcommand{\ud}{u^i\d_i}
\newcommand{\thd}{\th_{ia}d^{ia}}
\newcommand{\veth}{\vth^a\eth_a}

\newcommand{\icross}{\dlt^i_l\dlt^k_j}
\newcommand{\ipar}{\dlt^i_j\dlt^k_l}
\newcommand{\across}{\dlt^a_d\dlt^c_b}
\newcommand{\apar}{\dlt^a_b\dlt^c_d}

\newcommand{\CC}{{\mathbb C}}
\newcommand{\ZZ}{{\mathbb Z}}

\newcommand{\bal}{\bar\alpha}
\newcommand{\bbt}{\bar\beta}
\newcommand{\bgm}{\bar\gamma}
\newcommand{\bI}{\bar I}
\newcommand{\bJ}{\bar J}
\newcommand{\bZ}{\bar Z}

\title{{Structures Preserved by Consistently Graded Lie Superalgebras}}

\author{T. A. Larsson \\
Vanadisv\"agen 29, S-113 23 Stockholm, Sweden,\\
 email: Thomas.Larsson@hdd.se }

\maketitle
\begin{abstract}
Dual Pfaff equations (of the form $\tD^a = 0$, $\tD^a$ some vector fields 
of degree $-1$) preserved by the exceptional infinite-dimensional simple 
Lie superalgebras 
$\ksle(5|10)$, $\vle(3|6)$ and $\mb(3|8)$ are constructed, yielding
an intrinsic geometric definition of these algebras.
This leads to conditions on the vector fields, which are solved 
explicitly. Expressions for preserved differential form equations 
(Pfaff equations), brackets (similar to contact brackets)
and tensor modules are written down.
The analogous construction for the contact superalgebra $\kk(1|m)$
(a.k.a. the centerless $N=m$ superconformal algebra) is reviewed.
\end{abstract}

\newpage

\section{Introduction}
The simple Lie algebras of vector fields (v.f.) were classified by Cartan 
in 1909 \cite{Car09}. The classification consists of the four infinite 
series:

\renewcommand{\arraystretch}{1.1}

\begin{tabular}{ll}
$\vect(n)$ & general v.f. in $n$ dimensions,\\
$\svect(n)$ & divergence-free v.f.,\\
${\mathfrak h}(n)$ &Hamiltonian v.f ($n$ even),\\
$\kk(n)$ & contact v.f. ($n$ odd).
\end{tabular}

\noindent The corresponding problem in the super case was only recently 
settled \cite{CK99,Kac98,LSh00,Sh99}, building on earlier work in
\cite{ALS80,BL81,CK97,Kac77,Kac97,Lei85,LSh88,Lei90,Sh83,Sh97,ShP98}.
The classification consists of ten infinite series:

\begin{tabular}{ll}
$\vect(n|m)$
&{ arbitrary v.f. in $n|m$ dimensions, }\\
$\svect(n|m)$
&{ divergence-free v.f., }\\
${\mathfrak h}(n|m)$
&{ Hamiltonian v.f. ($n$ even), }\\
$\fle(n)$ 
&{ odd Hamiltonian or Leitesian v.f. $\subset \vect(n|n)$, }\\
$\sle(n)$ 
&{ divergence free Leitesian v.f., }\\
$\kk(n|m)$
&{ contact v.f. ($n$ odd), }\\
$\ko(n)$ 
&{ odd contact v.f. $\subset \vect(n|n+1)$, }\\
$\sko_\beta(n)$ 
&{ a deformation of div-free odd contact v.f.,} \\
$\widetilde\sle(n)$
&{ a deformation of $\sle(n)$}, \\
$\widetilde\sko(n)$
&{ a deformation of $\sko(n)$}.
\end{tabular}

A geometric way to describe these algebras is by stating what
structures they preserve, or what other conditions the vector fields obey:
\renewcommand{\arraystretch}{1.4}
\be
\begin{array}{|l|l|l|}
\hline
\hbox{Algebra} & \hbox{Basis} & \hbox{Description/structure preserved}\\
\hline
\vect(n|m) & u^i, \th^a & -\\
\svect(n|m) & u^i, \th^a & vol\\
{\mathfrak h}(n|m) & u^i, \th^a
& \om_{ij}du^idu^j + g_{ab}d\th^ad\th^b \\
\fle(n) & u^i, \th_i & du^id\th_i \\
\sle(n) & u^i, \th_i & du^id\th_i, vol \\
\kk(n+1|m) & t, u^i, \th^a
& dt + \om_{ij}u^idu^j + g_{ab}\th^ad\th^b = 0 \\
\ko(n) & \tau, u^i, \th_i
& d\tau + u^id\th_i + \th_idu^i = 0 \\
\sko_\beta(n)& \tau, u^i, \th_i
& M_f \in \ko(n): \div_\beta M_f = 0 \\
\widetilde\sle(n) & u^i, \th_i
& (1+\th_1..\th_n)X, X \in \sle(n)\\
\widetilde\sko(n) & \tau, u^i, \th_i
& (1+\th_1..\th_n)X, X \in \sko_{n+2\/n}(n)\\
\hline
\end{array}
\label{table}
\ee
\renewcommand{\arraystretch}{1.1}
In this table, $u^i$ denotes bosonic variables, $\th^a$ and $\th_i$
fermionic variables, and $t$ ($\tau$) is an extra bosonic (fermionic)
variable. The indices range over the dimensions indicated: $i=1,...,n$
and $a=1,...,m$. $\om_{ij} = -\om_{ji}$ and $g_{ab} = g_{ba}$ are
structure constants.
The notation $\al=0$ implies that it is this {\em Pfaff 
equation} that is preserved, not the form $\al$ itself; 
a vector field $X = X^\mu(x)\dmu$ acts on $\al$ as $\L_X\al = f_X\al$, 
$f_X$ some polynomial function.
$vol$ denotes the volume form; vector fields preserving $vol$ satisfy
$\div\, X \equiv (-)^{X\mu + \mu}\dmu X^\mu = 0$. 
\[
\div_\beta M_f = 2(-)^f ({\d^2f\/\d u^i\d\th_i} +
\big( u^i{\d\/\d u^i} + \th_i{\d\/\d \th_i} - n\beta){\d f\/\d\tau} \big)
\]
is a deformed divergence.

In addition, there exist five exceptional algebras, all discovered by 
Shchepochkina \cite{Sh83,Sh97,Sh99}: $\vas(4|4)$, $\ksle(5|10)$, 
$\vle(3|6)$, $\mb(3|8)$, and $\kas(1|6)$; the last case was 
independently found by Cheng and Kac \cite{CK97}. Here the numbers
indicate the super-dimension of the space on which the algebras are
realized. In fact, all these algebras except $\vas(4|4)$ have several
regradings, i.e. they can be realized on spaces of different 
super-dimensions. I will always
pick the {\em consistent} grading, i.e. the $\ZZ$-grading where
even subspaces are purely bosonic and odd subspaces purely fermionic.

In the mathematics literature, the exceptional Lie superalgebras are 
described in terms of Cartan prolongation. Unfortunately, this method
is defined 
recursively in a way which does not exhibit the geometric content.
The purpose of the present paper is to describe the structures preserved
by three of the four consistently graded exceptions. Conversely, 
knowledge of these
structures gives an intrinsic geometric definition of the algebras 
themselves. I also write down explicit equations satisfied by the vector
fields in these algebras, give general solutions to these equations, 
explicitly write down the brackets (analogous to Poisson or contact
brackets), and construct the tensor modules. 
This is listed as the first open problem in \cite{Sh99}.
The present paper also gives a partial answer to the third item in
Kac' vision list for the new millenium \cite{Kac99}.

Let $\oj \subset vect(n|m)$ be an algebra of polynomial vector fields 
acting on $\CC^{n|m}$. It has a Weisfeiler $\ZZ$-grading
of depth $d$ if it can be written as
\[
\oj = \oj_{-d} + ... +\oj_{-1} + \oj_0 + \oj_1 + ...,
\]
where the subspace $\oj_k$ consists of vector fields that are 
homogeneous of degree $k$, $\oj_0$ acts irreducibly on $\oj_{-1}$,
and $\oj_{-k} = \oj_{-1}^k$.
However, it is not the usual kind of homogeneity, because we do not 
assume that all directions are equivalent. Denote the coordinates of 
$n|m$-dimensional superspace by $x^\mu$ and let $\d_\mu$ be the 
corresponding derivatives. Then we define the grading by introducing 
positive integers $z_\mu$ such that $\deg x^\mu = z_\mu$ and 
$\deg \d_\mu = -z_\mu$.
The operator which computes the Weisfeiler grading is 
$Z = \sum_\mu z_\mu x^\mu \d_\mu$, and $\oj_k$ is the subspace of 
vector fields $X$ satisfying $[Z, X] = k X$. 
If we only considered $\oj$ as a graded vector space, we could of 
course make any choice of integers $z_\mu$, but we also want $\oj$ to be 
graded as a Lie algebra: $[\oj_i, \oj_j] = \oj_{i+j}$. 
The depth $d$ is identified with the maximal $z_\mu$.
We write $\oj_- =  \oj_{-d} + ... +\oj_{-1}$ and 
$\oj_+ = \oj_1 + \oj_2 + ...$; $\oj_-$ is a nilpotent superalgebra and
a $\oj_0$ module.

Cartan prolongation is defined as follows:
\begin{enumerate}
\item
Start with a realization for the non-positive part $\oj_0\ltimes\oj_-$
of $\oj$ in $n|m$-dimensional superspace.
\item
Define $\oj_k$ recursively for positive $k$ as the maximal subspace of 
$\vect(n|m)$ satisfying $[\oj_k, \oj_{-1}] \subset \oj_{k-1}$.
\end{enumerate}
In this paper I will use the following alternative method to construct 
the prolong:
\begin{enumerate}
\item
Again start with a realization for $\oj_0\ltimes\oj_-$.
\item
Determine the set of structures preserved by this algebra.
\item
Define the Cartan prolong 
$\oj = (\oj_{-d}, ..., \oj_{-1}, \oj_0)_* \equiv (\oj_-, \oj_0)_*,$
as the full subalgebra of $\vect(n|m)$ preserving the same structures.
\end{enumerate}
Clearly, the set of vector fields that preserve some structures 
automatically define a subalgebra of $\vect(n|m)$, so the problem is 
to find the right set. If the conditions imposed are too strong, the
solution may be a finite-dimensional algebra, and if they are too weak,
the resulting algebra may not be simple. In analogy with the known cases
listed above, it is natural to assume that such
a structure is either some differential form,
or an equation satisfied by forms (Pfaff equation), or a system of Pfaff
equations. 
However, although such Pfaff equations can be constructed
(and are so in this paper), it is simpler to consider the set of
{\em dual Pfaff equations}, which are of the form $\tD^a = 0$,
where $\tD^a$ is some vector field in $\vect(n|m)$ of degree $-1$
(necessarily not in 
$\oj$). Denote the space spanned by such $\tD^a$ by $\toj_{-1} \subset
\vect(n|m)$. We have $[Z,\tD^a]=-\tD^a$ and $[D^a,\tD^b] = 0$ for all 
$D^a\subset\oj_{-1}$. Since $\oj_{-k} = \oj_{-1}^k$, this implies
$[\toj_{-1}, \oj_{-k}] = 0$. Moreover, defining 
$\toj_{-k} = \toj_{-1}^k$ and $\toj_- = \bigoplus_k \toj_{-k}$, we see
that $[\toj_-, \oj_-] = 0$, so $\oj_-$ and $\toj_-$ form two commuting
subalgebras of $\vect(n|m)$. At least in all cases considered in this 
paper, these two subalgebras are isomorphic nilpotent algebras.
Now define the prolong 
$\oj= \{X\in\vect(n|m): [\tD^a, X] = f^a_b \tD^b\}$,
for each $\tD^a\in\toj_{-1}$, where $f^a_b$ are some polynomial functions,
depending on $X$.

To see that the second definition implies the first, let $D\in\oj_{-1}$
and $X\in\oj_k$. $Y=[D,X]$ satisfies $[\tD^a, Y] = [\tD^a, [D,X]]
= [D, [\tD^a,X]] = [D,f^a_b]\tD^b$, so $Y\in\oj_{k-1}$ and
$[\oj_{-1},\oj_k] \subset \oj_{k-1}$.

Vector fields that preserve the dual Pfaff equation $\toj_{-1}=0$ must
also preserve the higher order equations $\toj_{-2}=0$ and, for 
$\mb(3|8)$, $\toj_{-3} = 0$. This gives rise to further conditions obeyed
by the vector fields, but these additional relations are identities
which follow from $\toj_{-1}=0$. A new feature in the exceptions is
the appearance of certain symmetry conditions. These are genuinely
new constraints which have no counterpart in the contact algebra.

\renewcommand{\arraystretch}{1.4}

The algebras under consideration in this paper have the following 
description as Cartan prolongs:
\barr{l}
\ksle(5|10)
 = (\brep 5, \brep5^*\wwedge\brep5^*, sl(5))_*, \\
\vle(3|6)
 = (\brep 3\stimes\brep1, \brep3^*\stimes\brep2, \ssg)_*, \\
\mb(3|8) 
 = (\brep1\stimes\brep2, \brep 3\stimes\brep1, 
 \brep 3^*\stimes\brep 2, \ssg)_*, \\
\kas(1|6) \subset \kk(1|6) = (\brep 1, \brep 6, so(6)\ooplus gl(1))_*. 
\earr
\ee
Here $\stimes$ denotes the smash product, $\brep n$ is the
$n$-dimensional representation of $sl(n)$ or $so(n)$ and $\brep n^*$ its 
dual.
In addition, I apply the same technique to construct the contact algebras
$\kk(1|m) = (\brep 1, \brep m, so(m)\ooplus gl(1))_*$. Although the 
results for the contact algebras are not new, they are of interest for
several reasons:
\begin{enumerate}
\item
$\ksle(5|10)$, $\vle(3|6)$ and $\mb(3|8)$ are conceptually similar to 
$\kk(1|m)$, albeit more complicated, so this is a good place to develop
the machinery necessary for the exceptions.
\item
The exception $\kas(1|6)$ is a subalgebra of $\kk(1|6)$, so $\kas(1|6)$
satisfies the conditions described in section 2, together with some 
additional relations.
\item
$\kk(1|m)$ is consistently graded. In fact,
the only simple Lie superalgebras with consistent gradings are
$\kk(1|m)$, $\ksle(5|10)$, $\vle(3|6)$, $\mb(3|8)$ and $\kas(1|6)$
\cite{Kac98}. The present paper thus describes every consistently graded
simple Lie superalgebra, although no significant results are obtained for
$\kas(1|6)$.
\item
For small values of $m$, the Laurent polynomial version of $\kk(1|m)$
has a central extension, known in physics as the $N=m$ superconformal
algebra. The exceptional algebras do not admit central extensions, but
being subalgebras of $\vect(n|m)$ they clearly have non-central 
Virasoro-like extensions.
\end{enumerate}

For $\vle(3|6)$ and $\mb(3|8)$, the degree zero subalgebra $\oj_0 = \ssg$,
i.e. the non-compact form of the symmetries of the standard model in
particle physics. This suggests that these algebras may have important
applications to physics \cite{Kac99,Lar01}\footnote{
Note that the definition of $\mb(3|8)$ in \cite{Lar01} is flawed.}. 
To my knowledge, this is the only place where
the standard model algebra arises naturally and unambigously in a 
mathematically deep context. Note that $\oj_0$ is not just any subalgebra
of $\oj$, but that $\oj$ is completely determined by $\oj_0\ltimes\oj_-$.
Moreover, there is a 1-1 correspondence between $\oj_0$ and $\oj$ irreps. 
E.g., the $\vect(n) = (\brep n, gl(n))_*$ modules are tensor fields
and closed forms, corresponding to $gl(n)$ tensors.
Therefore, one may speculate that a $\oj$ symmetry may be mistaken 
experimentally for a $\oj_0$ symmetry.

Throughout this paper I use tensor calculus notation. $A^i$ denotes
a contravariant vector and $B_j$ a covariant vector. Repeated indices,
one up and one down, are implicitly summed over (Einstein convention). 
Derivatives are denoted by various types of d's ($d$, $\d$ and $\eth$).
$\eps^{ab}$, $\eps^{ijk}$, $\eps^{ijklm}$ and $\eps^{abcdef}$ denote
the totally anti-symmetric constant symbols in $\CC^2$, $\CC^3$, $\CC^5$
and $\CC^6$, respectively.
When dealing with the non-positive subalgebra $\oj_0\ltimes\oj_-$, all
parities are known and it is convenient to explicitly 
distinguish between anti-symmetric (straight) and symmetric (curly) 
brackets: $[A,B] = -[B,A]$ and $\{A,B\} = \{B,A\}$.
For general vector fields $X=X^\mu(x)\dmu$ (always assumed to be 
homogeneous in parity), the (straight) brackets are graded in the usual 
way: $[X,Y] = -(-)^{XY}[Y,X]$, where the symbol $(-)^X$ is $+1$ on 
bosonic components and $-1$ on fermionic ones. 
The sign convention is that $X$ acts as
\bes
\L_X\dnu &=& -(-)^{\nu X}\dnu X^\mu\dmu, \nle
\L_X dx^\mu &=& (-)^{(X+\mu+\nu)\nu} \dnu X^\mu dx^\nu.
\eens

The exceptional algebras are denoted by their names designed by 
Shchepochkina \cite{Sh99}. Kac and collaborators instead use the names
$E(5|10) = \ksle(5|10)$, $E(3|6) = \vle(3|6)$, $E(3|8) = \mb(3|8)$,
$E(1|6) = \kas(1|6)$, and $K(1|m) = \kk(1|m)$.
There are two reasons to choose Shchepochkina's convention. She
was the one who discovered the exceptions, and Kac' notation does not 
exhibit the family structure of regradings. E.g., $\mb(3|8)$ have three
regradings (realizations on superspaces of different dimensions):
$\mb(3|8)(4|5):2$, $\mb(3|8)(5|6):2$, $\mb(3|8):3$, where the number after
the colon indicates depth.

\section{$\kk(1|m)$}
Consider $\CC^{1|m}$ with basis spanned by one even coordinate $t$
and $m$ odd coordinates $\th_a$, $a=1,2,...,m$.
Let $\deg \th_a = 1$ and $\deg t = 2$.
The graded Heisenberg algebra has the non-zero relations
\be
\{d^a, \th_b\} = \dlt^a_b, \qquad [\d_0, t] = 1,
\ee
where $d^a = \d/\d\th_a$ and $\d_0 = \d/\d t$. Introduce an constant
metric $g_{ab} = g_{ba}$ with inverse $g^{ab}$. The metric and its 
inverse are used to raise and lower indices, so $\th^a = g^{ab}\th_b$
and $\th_a = g_{ab}\th^b$, etc.

The contact algebra $\kk(1|m)$ is generated by vector fields of the form
\be
K_f = (2-\th_ad^a)f\d_0 + (-)^fd_af d^a + \d_0 f\th_ad^a,
\label{Kf}
\ee
where $f = f(\th,t)$ is a function on $\CC^{1|m}$.
These vector fields satisfy the algebra $[K_f,K_g] = K_{[f,g]_K}$, where
the contact bracket reads
\be
[f,g]_K = (2-\th_ad^a)f\d_0g - \d_0f(2-\th_ad^a)g + (-)^fd_afd^ag.
\label{Kbrack}
\ee
By expanding $f(\th,t)$ in a power series in $\th_a$ we obtain more 
explicit descriptions of the contact bracket. For small $m=0,1,2,3,4$,
the resulting (Laurent polynomial) algebras are well known in physics, 
under the names centerless Virasoro and $N=m$ superconformal algebras, 
respectively. Note that (\ref{Kbrack}) is well defined also for $m\geq5$.

The non-positive part of $\kk(1|m)\subset\vect(1|m)$ is spanned by
the vector fields
\barr{|r|c|l|} 
\hline
\deg&f&\hbox{vector field}\\
\hline
-2& \half& E = \d_0 \\
-1& -\th_a& D^a = d^a - \th^a\d_0 \\
0& -\th^a\th^b & J^{ab} = \th^ad^b - \th^bd^a \\
0& t& Z = 2t\d_0 + \th_a d^a\\
\hline
\earr
\label{kfields}
\ee
The non-zero bracket in $\oj_-$ reads
\be
\{D^a, D^b\} = -2g^{ab}E.
\label{kcliff}
\ee
One notes that (\ref{kcliff}) defines a Clifford algebra, with $D^a$ 
playing the role of gamma matrices and $E$ that of the unit operator. 
The analogous nilpotent algebras for the exceptions, (\ref{kslecliff}), 
(\ref{vlecliff}) and (\ref{mbcliff}), constitute interesting 
generalizations of Clifford algebras.

A basis for $\toj_{-1}$ is given by
\be
\tD^a &=& d^a + \th^a\d_0 = D^a + 2\th^a\d_0.
\label{KD}
\ee
which satisfy
\bes
\{\tD^a, \tD^b\} &=& 2g^{ab}E, 
\nlb{kE}
\{D^a, \tD^b\} &=& 0.
\eens
Any vector field in $C^{1|m}$ has the form
\be
X = Q\d_0 + P_ad^a = \tQ\d_0 + P_a\tD^a,
\ee
where
\be
\tQ &=& Q + (-)^X\th^aP_a. 
\label{kQ}
\ee
$X$ preserves the dual Pfaff equation $\tD^a=0$, i.e.
\be
[X, \tD^a] = -(-)^X \tD^aP_b\tD^b,
\ee
provided that
\be
\tD^a\tQ = 2(-)^XP^a.
\label{keqn}
\ee
Compatibility between 
\be
[E, X] = \d_0\tQ E + \d_0P_a\tD^a
\ee
and (\ref{kE}) in the form $E = \half g_{ab}\{\tD^a, \tD^b\}$, implies that
\be
\d_0\tQ - {2\/m}(-)^X\tD^aP_a = 0.
\label{k2eqn}
\ee
However, this is an identity which follows from (\ref{keqn}) by
considering $\tD_a\tD^a\tQ$, so no new independent conditions
on the vector fields arise.

The pairing $\langle \tD^a, \al\rangle = 0$ gives
\be
\al = \dt + \th^a\dth_a,
\label{KPfaff}
\ee
or more explicitly $\al = \dt + g^{ab}\th_a\dth_b$.
We now show that $\al$ satisfies a Pfaff equation:
\bes
\L_X \th_a &=& P_a, \nl
\L_X t &=& Q, \nle
\L_X \dth_a &=& (-)^X d^bP_a\dth_b + \d_0P_a\dt, \nl
\L_X \dt&=& -(-)^X d^bQ\dth_b + \d_0Q \dt.
\eens
In particular,
\barr{lll}
E t = 1, &\ & E\th_a = E\dt = E\dth_a = 0, \\
D^a t = -\th^a, 
&&D^a\dt = -\dth^a, \\
D^a\th_b = \dlt^a_b,
&&D^a\dth_b = 0,\\
J^{ab} t = 0,
&&J^{ab}\dt = 0, \\
J^{ab}\th_c = \dlt^b_c\th^a - \dlt^a_c\th^b,
&&J^{ab}\dth_c = \dlt^b_c\dth^a - \dlt^a_c\dth^b, \\
Zt = 2t, 
&&Z\dt = 2\dt, \\
Z\th_a = \th_a, 
&&Z\dth_a = \dth_a.
\earr
\ee
One checks that
\be
J^{ab}\al = D^a\al = E\al = 0.
\ee
Thus the Pfaff equation $\al = 0$ is preserved by $\oj_-$
and $\oj_0$ and therefore by all of $\kk(1|m)\subset\vect(1|m)$.

An explicit calculation yields
\be
\L_X\al = \d_0\tQ\dt + (-(-)^Xd^b\tQ + 2P^b)\dth_b.
\ee
Since the Pfaff equation $\al=0$ is preserved, we have 
$\L_X\al = f\al = fdt + f\th^a\dth_a$ for $f = \d_0\tQ$. 
Substitution into the formula above shows that $X$ must satisfy 
(\ref{keqn}).

More generally, consider $\CC^{2n+1|m}$ with basis spanned by $2n$ even 
coordinates $u^i$, $i=1,2,...,2n$, $m$ odd coordinates $\th_a$, 
$a=1,2,...,m$, and one additional even coordinate $t$.
Let $\om_{ij} = -\om_{ji}$ be anti-symmetric structure constants and let
$g_{ab} = g_{ba}$, as before.
$\kk(2n+1|m)$ is the subalgebra of $\vect(2n+1|m)$ which
preserves the Pfaff equation $\al = 0$, where
\be
\al = \dt + \om_{ij}u^i\du^j + g^{ab}\th_a\dth_b.
\ee

Hence $\kk(1|m)$ has two natural classes of tensor modules, with bases
$\al$ and $\gm^a$ and module action
\bes
\L_X\al &=& \d_0\tQ\al, \nle
\L_X\gm^a &=& -(-)^X \tD^aP_b\gm^b.
\eens
Assuming that $\al$ and $\gm^a$ are fermions, we 
can now construct the volume form $v_\gm = \eps_{a_1a_2...a_m}
\gm^{a_1}\gm^{a_2}...\gm^{a_m }$, transforming as
\be
\L_Xv_\gm = -(-)^X \tD^aP_a v_\gm.
\ee
Since $\al$ has degree $+2$ and $v_\gm$ has degree $-m$, the form 
$v_\gm^2\al^2$ is invariant, which implies the relation (\ref{k2eqn}).
For $m = 2$, the invariant form is the volume form and (\ref{k2eqn}) 
becomes $\div\, X = 0$.

{F}rom the vector density $\gm^a$ with weight $-1$ we can construct a 
vector $\bgm^a$ of zero weight, transforming as 
\bes
\L_X\bgm^a &=& -(-)^X(\tD^aP_b - {1\/m}\dlt^a_b\tD^cP_c)\bgm^b \nle
&=& \fourth \tD_b\tD_c\tQ(g^{ac}\bgm^b - g^{ab}\bgm^c).
\eens
Hence we obtain the explicit realization
\be
\L_X = X + \fourth \tD_a\tD_b\tQ\bJ^{ab} + \half\d_0\tQ\bZ,
\label{ktensor}
\ee
where $X$ acts trivially on $\bgm^a$ and
\bes
\bJ^{ab}\bgm^c &=& g^{ac}\bgm^b - g^{bc}\bgm^a, \nle
\bZ\bgm^c &=& 0.
\eens
Thus $\bJ^{ab}$ and $\bZ$ generate the Lie algebra 
$\boj_0 = so(m)\ooplus gl(1)$. However, it is clear that $\L_X$ 
will satisfy the same algebra for every representation of $\boj_0$.
Substitution of irreducible $\boj_0$ modules into (\ref{ktensor}) 
gives the tensor modules for $\kk(1|m)$. In particular, the expressions
for $\oj_0$ becomes $\L_{J^{ab}} = J^{ab}+\bJ^{ab}$, $\L_Z = Z + \bZ$,
i.e. two commuting copies of $\oj_0$.

To obtain an explicit expression for the vector fields in $\kk(1|m)$,
we set $\tQ = f(\th,u)$, an arbitrary polynomial function. {F}rom
(\ref{kQ}) and (\ref{keqn}) we obtain $P^a = \mhalf \tD^af$, and
\be
X = \half K_f = f\d_0 + \half(-)^f \tD^af\tD_a.
\ee 
We have $[X_f, X_g] = X_{[f,g]_{\kk}}$, where
\be
[f,g]_{\kk} = f\d_0g - (-)^{fg}g\d_0f + \half(-)^f \tD^af\tD_ag.
\ee
One checks that this contact bracket is related to  (\ref{Kbrack}) by
$[f,g]_{\kk} = \half [f,g]_K$.

\section{$\kas(1|6)$}
$\kk(1|6)$ contains the exceptional simple subalgebra $\kas(1|6)$. 
Clearly every vector field in $\kas(1|6)$ preserves the Pfaff equation
({\ref{KPfaff}), but in addition there is a condition coming from 
degree $+1$. The description here closely follows \cite{CK99}.

Let $j^{ab} = -\th^a\th^b$, so $K_{j^{ab}} = J^{ab}$ generate $so(m)$.
We have $[j^{ab}, \th^c]_K = g^{bc}\th^a - g^{ac}\th^b$ and
\be
[j^{ab}, \th^c\th^d\th^e]_K = g^{bc}\th^a\th^d\th^e
- g^{ac}\th^b\th^d\th^e + \hbox{4 cyclic terms}.
\ee
Therefore, 
$\omega^{abc}_+ = \th^a\th^b\th^c + \eps^{abcdef}\th_d\th_e\th_f$
and 
$\omega^{abc}_- = \th^a\th^b\th^c - \eps^{abcdef}\th_d\th_e\th_f$
transform independently under $so(6)$; recall that the metric 
$g_{ab}$ is used to lower indices. Denote the corresponding
$so(6)$ modules $V_+$ and $V_-$, respectively, and let $M$ be the
$so(6)$ module corresponding to $f = t\th^a$.
$\kas(1|6)$ is obtained by requiring that $\oj_1 = M\ooplus V_+$
(or equivalently $\oj_1 = M\ooplus V_-$). In contrast,  
$\oj_1 = M\ooplus V_+\ooplus V_-$ for the contact algebra $\kk(1|6)$.

The contact vector fields in $\kk(1|m)$ at degree $+1$ are
\barr{|r|c|l|}
\hline
\deg&f&K_f\\
\hline
1 & t\th_a& A_a = t\th_a\d_0 - td_a + \th_a\th_bd^b \\
1 & -\th_a\th_b\th_c & B_{abc} = \th_a\th_b\th_c\d_0 
+ \th_a\th_bd_c + \th_b\th_cd_a + \th_c\th_ad_b\\
\hline
\earr
\label{kasfields}
\ee
For $m=6$, we can define the dual of $B_{abc}$ as
\be
B^*_{abc} = \eps_{abcdef}B^{def}.
\ee
$\kas(1|6)$ is obtained by requiring that $B_{abc}$ be self-dual, 
i.e. $B_{abc} = B^*_{abc}$. 

A similar construction can be carried out for $m=4$; the corresponding
subalgebra of $\kk(1|4)$ is finite-dimensional \cite{CK99}. 
The $\kk(1|2n)$ subspace $\oj_n$ contains two subspaces that are 
isomorphic as $so(2n)$ modules, but it is not possible to eliminate 
one of them when $2n > 6$.

I have not been able to formulate $\kas(1|6)$ as an algebra of vector
fields preserving some Pfaff equation or dual Pfaff equation.

\section{$\ksle(5|10)$}
Let $u = (u^i)\in\CC^5$, $\d_i = \d/\d u^i$, $i= 1,2,3,4,5$.
Let $\xi = \xi^i(u)\d_i$ be a divergence-free vector field: 
$\d_i \xi^i = 0$,
and let $\om = \om_{ij}(u)du^i\wwedge du^j$ be a closed two-form:
$\om_{ji} = -\om_{ij}$ and 
$\d_i \om_{jk} +\d_j \om_{ki} +\d_k \om_{ij} = 0$.
$\ksle(5|10)$  has generators $\L_\xi=\L_i(\xi^i)$ and 
$\G_\om=\G^{ij}(\om_{ij})$. The brackets read \cite{CK99}
\bes
[\L_\xi,\L_\eta] &=& \L_{[\xi,\eta]}
= \L_k(\xi^i\d_i\yk - \yj\d_j\xi^k), \nl
{[}\L_\xi, \G_\om] &=& 
 \G^{jk}(\xi^i\d_i\om_{jk} + \d_j\xi^i\om_{ik} + \d_k\xi^i\om_{ji}), \\
\{\G_\om, \G_\ups\} &=&
  \eps^{ijklm} \L_m( \om_{ij} \ups_{kl}).
\eens

Consider $\CC^{5|10}$ with basis spanned by five even coordinates
$u^i$, $i=1,2,3,4,5$ and ten odd coordinates $\th_{ij} = -\th_{ji}$.
Let $\deg \th_{ij} = 1$ and $\deg u^i = 2$.
The graded Heisenberg algebra has the non-zero relations
\be
[\d_j, u^i] = \dlt^i_j, \qquad \{d^{ij}, \th_{kl}\} 
= \dlt^i_k\dlt^j_l - \dlt^j_k\dlt^i_l,
\ee
where $\d_i = \d/\d u^i$ and $d^{ij} = -d^{ji} = \d/\d\th_{ij}$.
The non-positive part of $\ksle(5|10)\subset\vect(5|10)$ is spanned by
the vector fields
\barr{|r|l|} 
\hline
\deg&\hbox{vector field}\\
\hline
-2& E_k = \d_k \\
-1& D^{kl} = d^{kl} -\half \eps^{klmni}\th_{mn}\d_i \\
0& I^k_l = u^k\d_l - \th_{lj} d^{kj} 
- \fifth\dlt^k_l(\ud - \th_{ij} d^{ij}) \\
\hline
\earr
\label{kslefields}
\ee
Note that $Z=2u^i\d_i + \half\th_{ij}d^{ij}$, which is the operator that
computes the Weisfeiler grading, is not part of $\oj_0$. However, for 
convenience we will also consider the non-simple algebra 
$\bigksle = (\brep 5, \brep5^*\wwedge\brep5^*, gl(5))_*$, obtained 
by adjoining $Z$ to (\ref{kslefields}). It was shown in \cite{CK99}
that $\bigksle = \ksle(5|10) \ooplus \CC Z$.

The non-zero brackets in $\oj_-$ read
\be
\{D^{ij}, D^{kl}\} = -2\eps^{ijklm}E_m.
\label{kslecliff}
\ee
A basis for $\toj_{-1}$ is given by
\be
\tD^{ij} &=& d^{ij} + \half\eps^{ijklm}\th_{kl}\d_m
= D^{ij} + \eps^{ijklm}\th_{kl}\d_m, 
\label{ksleD}
\ee
which satisfy
\bes
\{\tD^{ij}, \tD^{kl}\} &=& 2\eps^{ijklm}E_m, 
\nlb{ksleE}
\{D^{ij}, \tD^{kl}\} &=& 0.
\eens
Any vector field in $\vect(5|10)$ has the form
\be
X = Q^i\d_i + \half P_{ij}d^{ij} = \tQ^i\d_i + \half P_{ij}\tD^{ij},
\ee
where the $\half$ is necessary to avoid double counting and
\be
\tQ^i &=& Q^i + \mfourth\eps^{ijklm}\th_{jk}P_{lm}.
\label{ksleQ}
\ee
$X$ preserves the dual Pfaff equation $\tD^{ij}=0$, i.e.
\be
[X, \tD^{ij}] = -\mhalf \tD^{ij}P_{kl}\tD^{kl},
\ee
provided that
\be
\tD^{ij}\tQ^k = (-)^X\eps^{ijklm}P_{lm}.
\label{ksleeqn}
\ee
In particular, we have the symmetry relations
\be
\tD^{ij}\tQ^k = \tD^{jk}\tQ^i = -\tD^{ik}\tQ^j.
\label{kslesymm}
\ee
Compatibility between 
\be
[E_i, X] = \d_i\tQ^j E_j + \half\d_iP_{jk}\tD^{jk}
\ee
and (\ref{ksleE}) in the form 
$E_i = {1\/48}\eps_{ijklm}\{\tD^{jk}, \tD^{lm}\}$, implies that
\be
\d_i\tQ^j = \msixth(\tD^{kl}P_{kl}\dlt^j_i - 2\tD^{jk}P_{ik}).
\label{ksle2eqn}
\ee
In particular,
\be
\d_i\tQ^i -\mhalf \tD^{ij}P_{ij} \equiv \div\, X = 0.
\label{kslediv}
\ee
However, (\ref{ksle2eqn}) is an identity which follows from
(\ref{ksleeqn}) by considering 
$\eps_{ijkln}\tD^{ij}\tD^{kl}\tQ^m$, so no new independent 
conditions on the vector fields arise.

The pairing $\langle \tD^{ij}, \al^k\rangle = 0$ gives
\be
\al^i = \du^i + \fourth\eps^{ijklm}\th_{jk}\dth_{lm}.
\ee
We now show that $\al^i$ satisfies a Pfaff equation:
\bes
\L_X \th_{ij} &=& P_{ij}, \nl
\L_X u^i &=& Q^i, \nle
\L_X \dth_{ij} &=& \mhalf d^{kl}P_{ij}\dth_{kl} 
+ \d_kP_{ij}\du^k, \nl
\L_X \du^i &=& -\mhalf d^{kl}Q^i\dth_{kl} + \d_kQ^i\du^k.
\eens
In particular,
\barr{lll}
E_k u^i = \dlt^i_k,
&& E_k\du^i = E_k \th_{ij} = E_k\dth_{ij} = 0, \\
D^{kl}u^i = -\half\eps^{klmni}\th_{mn}, 
&\ &D^{kl}\du^i = -\half\eps^{klmni}\dth_{mn}, \\
D^{kl}\th_{ij} = \dlt^k_i\dlt^l_j - \dlt^l_i\dlt^k_j, 
&&D^{kl}\dth_{ij} = 0, \\
I^k_lu^i = \dlt^i_l u^k - \fifth \dlt^k_l u^i, 
&&I^k_l\du^i = \dlt^i_l \du^k - \fifth \dlt^k_l \du^i, \\
I^k_l\th_{ij} = -\dlt^k_i\th_{lj} - \dlt^k_j\th_{il}
+ {2\/5}\dlt^k_l\th_{ij}, 
&&I^k_l\dth_{ij} = -\dlt^k_i\dth_{lj} - \dlt^k_j\dth_{il}
+ {2\/5}\dlt^k_l\dth_{ij},\\
Zu^i = 2u^i, && Z\th_{ij} = \th_{ij}, \\
Z\du^i = 2\du^i, && Z\dth_{ij} = \dth_{ij}.
\earr
\ee
One checks that
\bes
&&I^k_l\al^i = \dlt^i_l \al^k - \fifth \dlt^k_l \al^i, \nle
&&D^{kl}\al^i = E_k\al^i = 0.
\eens
An explicit calculation yields
\be
\L_X\al^i = \d_j\tQ^i\du^j + 
(-\mhalf d^{np}\tQ^i + \half\eps^{ijknp}P_{jk})\dth_{np}.
\ee
Since the Pfaff equation $\al^i=0$ is preserved, we have 
$\L_X\al^i = f^i_j\al^j = 
f^i_j(\du^j + \fourth\eps^{jklmn}\th_{kl}\dth_{mn})$ for 
$f^i_j = \d_j\tQ^i$. 
Substitution into the formula above yields the condition
(\ref{ksleeqn}).

Hence $\ksle(5|10)$ has two natural classes of tensor modules, 
with bases $\al^i$ and $\gm^{ij}$ and module action
\bes
\L_X\al^i &=& \d_j\tQ^i\al^j, 
\nlb{kslemod}
\L_X\gm^{ij} &=& -\mhalf \tD^{ij}P_{kl}\gm^{kl}.
\eens
Assuming that $\al^i$ and $\gm^{ij}$ are fermions, we 
can now construct the volume forms
$v_\al = \al^1\al^2\al^3\al^4\al^5$ and
$v_\gm = \gm^{12}\gm^{13}\gm^{14}\gm^{15}\gm^{23}
\gm^{24}\gm^{25}\gm^{34}\gm^{35}\gm^{45}$, transforming as
\bes
\L_Xv_\al &=& \d_i\tQ^iv_\al, \nle
\L_Xv_\gm &=& -\mhalf \tD^{ij}P_{ij} v_\gm.
\eens
Since $v_\al$ has degree $+10$ and $v_\gm$ has degree $-10$, the form 
$v_\al v_\gm$ is invariant, which implies the relations (\ref{kslediv}).

{F}rom (\ref{kslemod}) we deduce the transformation laws for a scalar
density $v$ of weight $+1$, an $sl(5)$ vector $\bal^i$ and an 
$sl(5)$ bivector $\bgm^{ij}$:
\bes
\L_Xv &=& \hbox{$1\/10$}\d_i\tQ^i v 
= \hbox{$1\/20$}(-)^X\tD^{ij}P_{ij}v, \nl
\L_X\bal^i &=& 
\hbox{$1\/15$}(\dlt^i_j\tD^{kl}P_{kl} - 5\tD^{ik}P_{jk})\bal^j\nle
&=& (\d_j\tQ^i - \fifth\dlt^i_j\d_k\tQ^k)\bal^j, \nl
\L_X\bgm^{ij} &=& \hbox{$1\/20$}(\dlt^i_k\dlt^j_l\tD^{mn}P_{mn} 
-10\tD^{ij}P_{kl})\bgm^{kl}.
\eens
Hence we obtain the explicit realization
\bes
\L_X &=& X - \mthird \tD^{ik}P_{jk}\bI^j_i 
+ \hbox{$1\/20$}(-)^X\tD^{ij}P_{ij}\bZ
\nlb{ksletensor}
&=& X + \d_j\tQ^i\bI^j_i + \hbox{$1\/10$}\d_i\tQ^i \bZ,
\eens
where $\bI^i_j$ and $\bZ$ generate the Lie algebra $\boj_0 = gl(5)$ 
and $X$ commutes with $\boj_0$. 
This expression holds for the non-simple algebra $\bigksle$;
for $\ksle(5|10)$ $\bZ=0$.
By introducing canonical conjugate
oscillators $\bal^*_i$ and $v^*$, subject to the Heisenberg algebra
$[\bal^*_i, \bal^j] = \dlt^j_i$, $[v^*, v] = 1$, we obtain the explicit
expression for the $\boj_0$ generators: 
$\bI^i_j = \bal^i\bal^*_j - \fifth\dlt^i_j\bal^k\bal^*_k$ and
$\bZ = vv^*$. However, it is clear that $\L_X$ 
will satisfy the same algebra for every representation of $gl(5)$.
Substitution of irreducible $\boj_0$ modules into (\ref{ksletensor}) 
gives the tensor modules for $\bigksle$. In particular, the 
expressions for $\oj_0$ become $\L_{I^i_j} = I^i_j + \bI^i_j$, 
$\L_Z = Z + \bZ$, i.e. two commuting copies of $\oj_0$.

To obtain an explicit expression for the vector fields in $\ksle(5|10)$,
we set $\tQ^i = f^i(\th,u)$, five arbitrary polynomial 
functions. {F}rom (\ref{ksleQ}) and (\ref{ksleeqn}) we obtain
$P_{ij} = {1\/12}(-)^X\eps_{ijklm} \tD^{kl}f^m$ and
\be
X = U_f = f^i\d_i + \hbox{$1\/24$}(-)^f\eps_{ijklm} \tD^{ij}f^k\tD^{lm},
\ee
where $f = f^i(\th,u)\d_i$ is a vector field acting on $\CC^{5|10}$ and 
$(-)^f = +1$ if $f$ is an even vector field, i.e. $f^i$ is an even
function.
Due to the symmetry condition (\ref{kslesymm}), the 
components $f^i$ are not independent, but subject to the relation
\bes
\tD^{ij}f^k = \tD^{jk}f^i = -\tD^{ik}f^j, 
\label{kslefsymm}
\ee
This condition may alternatively be written as
$2\tD^{ij}f^k = \tD^{jk}f^i + \tD^{ki}f^j$.

There is one additional relation, which singles out $\ksle(5|10)\subset
\bigksle$:
\be
\d_i\tQ^i = 0 \Rightarrow \d_if^i = 0.
\label{divQ}
\ee
This extra condition is clearly satisfied by all operators in 
(\ref{kslefields}), except for $Z$.

A $\ksle(5|10)$ bracket is defined by $[U_f, U_g] = U_{[f,g]_{\ksle}}$,
where
\bes
[f,g]_{\ksle} &=& f^i\d_ig^j\d_j - \d_jf^ig^j\d_i 
+ (-)^f H^p_{ijk|lmn}\tD^{ij}f^k\tD^{lm}g^n\d_p	\nl
&&- (-)^{g+fg} H^p_{lmn|ijk}\tD^{lm}g^n\tD^{ij}f^k\d_p, \\
H^p_{ijk|lmn} &=& \hbox{$1\/24$}\eps_{ijklm}\dlt^p_n
-\hbox{$1\/288$}\eps_{ijkqr}\eps_{lmnst}\eps^{qrstp}.
\eens
Finally we list the vector fields $f$ and $U_f$ for $\oj_0\ltimes\oj_-$:
\barr{|r|l|l|} 
\hline
\deg&f&U_f\\
\hline
-2& \d_i & E_i \\
-1& -\eps^{ijklm}\th_{kl}\d_m & D^{ij} \\
0& u^i\d_j - \fifth \dlt^i_ju^k\d_k 
+ \half\eps^{iklmn}\th_{jk}\th_{lm}\d_n& I^i_j\\
0 & 2u^i\d_i & Z \\
\hline
\earr
\ee
Other vector fields $f$ of non-positive degree are not compatible with the 
symmetry condition (\ref{kslefsymm}). Clearly, the vector field 
corresponding to $U_f=Z$ does not satisfy (\ref{divQ}).

\section{$\vle(3|6)$}
Let $u = (u^i)\in\CC^3$, $i= 1,2,3$.
Let $\xi = \xi^i(u)\d/\d u^i$ be a vector field,
let $X = X^a_b(u) T^b_a$ be an $sl(2)$-valued current,
and let $\om = \om_{ia}(u)\du^i\dz^a$ be a $\CC^2$-valued one-form.
Here $T^a_b \in sl(2)$:
$[T^a_b, T^c_d] = \dlt^c_b T^a_d - \dlt^a_d T^c_b$,
and $X^a_a = 0$.
$\vle(3|6)$ has generators $\L_\xi=\L_i(\xi^i)$, $\J_X=J^b_a(X^a_b)$ 
and $\G_\om=\G^{ia}(\om_{ia})$, and the brackets read \cite{CK99}
\bes
[\L_\xi,\L_\eta] &=& \L_{[\xi,\eta]}
= \L_k(\xi^i\d_i\yk - \yj\d_j\xi^k), \nl
{[}\L_\xi,\J_X] &=& \J_{\xi X} = \J^b_a(\xi^i\d_i X^a_b),\nl
{[}\J_X,\J_Y] &=& \J_{[X,Y]} = \J^b_a(Y^a_cX^c_b - X^a_cY^c_b),\nle
{[}\L_\xi, \G_\om] &=& \G^{ja}(\xi^i\d_i\om_{ja} 
+ \d_j\xi^i\om_{ia} + \half\d_i\xi^i\om_{ja}), \nl
{[}\J_X, \G_\om] &=&  \G^{ia}( X^b_a\om_{ib}), \nl
\{\G_\om, \G_\ups\} &=&
 \eps^{ijk}\eps^{ab} \L_k( \om_{ia} \ups_{jb})
 + \eps^{ijk}\eps^{ac} \J^b_a(\d_k\om_{ic} \ups_{jb}
 -\om_{ib} \d_k\ups_{jc}).
\eens

In this and the next section we use some special relations valid in
two dimensions only:
\bes
&&\phi^a = \eps^{ab}\phi_b, \qquad \phi_a = \eps_{ab}\phi^b, 
\label{raise}\\
&&\eps^{ab}\eps_{bc} = \dlt^a_c, \qquad 
\eps^{ab}\eps_{cb} = -\dlt^a_c, \\
&&\phi^a\psi_a = -\phi_a\psi^a, \\
&&\eps^{ab}\phi^c + \eps^{bc}\phi^a + \eps^{ca}\phi^b = 0,
\label{ident}\\
&&\phi^a\psi^b - \phi^b\psi^a = -\eps^{ab}\phi^c\psi_c.
\label{skew}
\ees
Our convention is $\eps^{12} = \eps_{21} = +1$, 
$\eps^{21} = \eps_{12} = -1$. The constants $\eps^{ab}$ and $\eps_{ab}$ 
can be used to raise and lower $sl(2)$ indices. 

Consider $\CC^{3|6}$ with basis spanned by three even coordinates
$u^i$, $i=1,2,3$ and six odd coordinates $\th_{ia}$.
Let $\deg \th_{ia} = 1$ and $\deg u^i = 2$.
The graded Heisenberg algebra has the non-zero relations
\be
[\d_j, u^i] = \dlt^i_j, \qquad 
\{d^{ia}, \th_{jb}\} = \dlt^i_j\dlt^a_b,
\ee
where $\d_i = \d/\d u^i$ and $d^{ia} = \d/\d\th_{ia}$.
The non-positive part of $\vle(3|6)\subset\vect(3|6)$ is spanned by
the vector fields
\barr{|r|l|}
\hline
\deg&\hbox{vector field} \\
\hline
-2& E_i = \d_i \\
-1& D^{ia} = d^{ia} - \eps^{ijk}\th^a_j\d_k \\
0& I^k_l = u^k\d_l - \th_{la} d^{ka} 
- \third\dlt^k_l(\ud - \thd) \\
0& J^c_d = -\th_{id} d^{ic} + \half\dlt^c_d\thd \\
0& Z = 2u^i\d_i + \th_{ia}d^{ia} \\
\hline
\earr
\label{vlefields}
\ee
The non-zero brackets in $\oj_-$ are
\be
\{D^{ia}, D^{jb}\} = -2\eps^{ijk}\eps^{ab}E_k.
\label{vlecliff}
\ee
A basis for $\toj_{-1}$ is given by
\be
\tD^{ia} &=& d^{ia} + \eps^{ijk}\th^a_j\d_k
= D^{ia} + 2\eps^{ijk}\th^a_j\d_k, 
\label{vleD}
\ee
which satisfy
\bes
\{\tD^{ia}, \tD^{jb}\} &=& 2\eps^{ijk}\eps^{ab}E_k, 
\nlb{vleE}
\{D^{ia}, \tD^{jb}\} &=& 0.
\eens
Any vector field in $\vect(3|6)$ has the form
\be
X = Q^i\d_i + P_{ia}d^{ia} = \tQ^i\d_i + P_{ia}\tD^{ia},
\ee
where
\be
\tQ^i &=& Q^i - (-)^X\eps^{ijk}\th^a_jP_{ka}. \nl
\label{vleQ}
\ee
$X$ preserves the dual Pfaff equation $\tD^{ia}=0$, i.e.
\be
[X, \tD^{ia}] = -(-)^X \tD^{ia}P_{jb}\tD^{jb},
\ee
provided that
\be
\tD^{ia}\tQ^j = -2(-)^X\eps^{ijk}P^a_k.
\label{vleeqn}
\ee
In particular, we have the symmetry relations
\be
\tD^{ia}\tQ^j = -\tD^{ja}\tQ^i.
\label{vlesymm}
\ee
Compatibility between 
\be
[E_i, X] = \d_i\tQ^j E_j + \d_iP_{jb}\tD^{jb}
\ee
and (\ref{vleE}) in the form 
$E_i = {1\/8}\eps_{ijk}\eps_{ab}\{\tD^{ja}, \tD^{kb}\}$, implies
that
\be
\d_i\tQ^j = \mhalf(\tD^{ka}P_{ka}\dlt^j_i - \tD^{ja}P_{ia}).
\label{vle2eqn}
\ee
In particular,
\be
\d_i\tQ^i  - (-)^X \tD^{ia}P_{ia} \equiv \div\, X = 0.
\label{vlediv}
\ee
However, (\ref{vle2eqn}) is an identity which follows from
(\ref{vleeqn}) by considering 
$\eps_{ijl}\eps_{ab}\tD^{ia}\tD^{jb}\tQ^k$, so no new independent 
conditions on the vector fields arise.

The pairing $\langle \tD^{ia}, \al^j\rangle = 0$ gives
\be
\al^i = \du^i - \eps^{ijk}\th^a_j\dth_{ka}.
\label{vlePfaff}
\ee
We now show that $\al^i$ satisfies a Pfaff equation:
\bes
\L_X \th_{ia} &=& P_{ia}, \nl
\L_X u^i &=& Q^i, \nle
\L_X \dth_{ia} &=& (-)^X d^{jb}P_{ia}\dth_{jb} + \d_jP_{ia}\du^j, \nl
\L_X \du^i &=& -(-)^X d^{jb}Q^i\dth_{jb} + \d_jQ^i\du^j.
\eens
In particular,
\barr{lll}
E_k u^i = \dlt^i_k, &\ & E_k\du^i = E_k\th_{ia} = E_k\dth_{ia} = 0,\\
D^{kc}u^i = -\eps^{kli}\th^c_l, 
&&D^{kc}\du^i = -\eps^{kli}\dth^c_l, \\
D^{kc}\th_{ia} = \dlt^k_i\dlt^c_a, 
&&D^{kl}\dth_{ia} = 0, \\
I^k_lu^i = \dlt^i_l u^k - \third \dlt^k_l u^i,
&&I^k_l\du^i = \dlt^i_l \du^k - \third \dlt^k_l \du^i, \\
I^k_l\th_{ia} = -\dlt^k_i\th_{la} + \third\dlt^k_l\th_{ia}, 
&&I^k_l\dth_{ia} = -\dlt^k_i\dth_{la} + \third\dlt^k_l\dth_{ia},\\
J^c_d u^i = 0, 
&&J^c_d \du^i = 0, \\
J^c_d\th_{ia} = -\dlt^c_a\th_{id} + \half\dlt^c_d\th_{ia}, 
&&J^c_d\dth_{ia} = -\dlt^c_a\dth_{id} + \half\dlt^c_d\dth_{ia}, \\
Zu^i = 2u^i, 
&&Z\du^i = 2\du^i, \\
Z\th_{ia} = \th_{ia}, 
&& Z\dth_{ia} = \dth_{ia}.
\earr
\ee
One checks that
\bes
&&I^k_l\al^i = \dlt^i_l \al^k - \third \dlt^k_l \al^i, \nl
&&Z\al^i = 2\al^i, \\
&&J^a_b\al^i = D^{kl}\al^i = E_k\al_i = 0.
\eens
An explicit calculation yields
\be
\L_X\al^i = \d_j\tQ^i\du^j + 
(-(-)^X d^{lb}\tQ^i - 2\eps^{lij}P^b_j)\dth_{lb}.
\ee
Since the Pfaff equation $\al^i=0$ is preserved, we have 
$\L_X\al^i = f^i_j\al^j = 
f^i_j(\du^j - \eps^{jkl}\th^a_k\dth_{la})$ for 
$f^i_j = \d_j\tQ^i$. 
Substitution into the formula above yields the conditions (\ref{vleeqn}).

Hence $\vle(3|6)$ has two natural classes of tensor modules, 
with bases $\al^i$ and $\gm^{ia}$ and module action
\bes
\L_X\al^i &=& \d_j\tQ^i\al^j, \nle
\L_X\gm^{ia} &=& -(-)^X \tD^{ia}P_{jb}\gm^{jb}.
\eens
Assuming that $\al^i$ and $\gm^{ia}$ are fermions, we 
can now construct the volume forms
$v_\al = \eps_{ijk}\al^i\al^j\al^k$ and \allowbreak
$v_\gm = \eps_{ijk}\eps_{lmn}\eps_{ad}\eps_{be}\eps_{cf}
\gm^{ia}\gm^{jb}\gm^{kc}\gm^{ld}\gm^{me}\gm^{nf}$, transforming as
\bes
\L_Xv_\al &=& \d_i\tQ^iv_\al, 
\nlb{vlemod}
\L_Xv_\gm &=& -(-)^X \tD^{ia}P_{ia} v_\gm.
\eens
Since $v_\al$ has degree $+6$ and $v_\gm$ has degree $-6$, the form 
$v_\al v_\gm$ is invariant, which implies the relation (\ref{vlediv}).

{F}rom (\ref{vlemod}) we deduce the transformation laws for a scalar 
density $v$ of weight $+1$, an $sl(3)$ vector $\bal^i$ and an 
$sl(3)\ooplus sl(2)$ vector $\bgm^{ia}$:
\bes
\L_Xv &=& \sixth\d_i\tQ^i v 
= \msixth\tD^{ia}P_{ia}v, \nl
\L_X\bal^i &=& 
\msixth(\dlt^i_j\tD^{ka}P_{ka} - 3\tD^{ia}P_{ja})\bal^j
\nlb{vlemmod}
&=& (\d_j\tQ^i - \third\dlt^i_j\d_k\tQ^k)\bal^j,\nl
\L_X\bgm^{ia} &=& \msixth(\dlt^i_j\dlt^a_b\tD^{kc}P_{kc}
-6\tD^{ia}P_{jb})\bgm^{jb}.
\eens
I have not obtained any explicit expression for an $sl(2)$ vector 
$\bbt^a$, but in view of (\ref{vlemmod}) it is natural to assume that 
it transforms as
\be
\L_X\bbt^a = \msixth(\dlt^a_b\tD^{ic}P_{ic} - 2\tD^{ia}P_{ib})\bbt^b.
\label{vleguess}
\ee
Hence we obtain the explicit realization
\bes
\L_X &=& X + \msixth \tD^{ia}P_{jb}(-3\dlt^b_a\bI^j_i - 2\dlt^j_i\bJ^b_a
+ \dlt^j_i\dlt^b_a\bZ) 
\nlb{vletensor}
&=& X + \d_j\tQ^i\bI^j_i 
- \mthird\tD^{ia}P_{ib}\bJ^b_a+ \sixth\d_i\tQ^i \bZ,
\eens
where $\bI^i_j$, $\bJ^a_b$ and $\bZ$ generate the Lie algebra 
$\boj_0 = \ssg$ and $X$ commutes with $\boj_0$. 
Note that the term multiplying $\bJ^b_a$ is somewhat uncertain, since
(\ref{vleguess}) is a conjecture.
By introducing canonical conjugate oscillators $\bal^*_i$, $\bbt^*_a$
and $v^*$, subject to the Heisenberg algebra
$[\bal^*_i, \bal^j] = \dlt^j_i$, $[\bbt^*_a, \bbt^b] = \dlt^b_a$, 
$[v^*, v] = 1$, we obtain the explicit expression for the $\boj_0$ 
generators: 
$\bI^i_j = \bal^i\bal^*_j - \third\dlt^i_j\bal^k\bal^*_k$, 
$\bJ^a_b = \bbt^a\bbt^*_b - \half\dlt^a_b\bbt^c\bbt^*_c$ and
$\bZ = vv^*$. However, it is clear that $\L_X$ 
will satisfy the same algebra for every representation of $\ssg$.
Substitution of irreducible $\boj_0$ modules into (\ref{vletensor}) 
gives the tensor modules for $\vle(3|6)$. In particular, the expressions
for $\oj_0$ become $\L_{I^i_j} = I^i_j + \bI^i_j$,
$\L_{J^a_b} = J^a_b + \bJ^a_b$ and $\L_Z = Z + \bZ$,
i.e. two commuting copies of $\oj_0$.

To obtain an explicit expression for the vector fields in $\vle(3|6)$,
we set $\tQ^i = f^i(\th,u)$, three arbitrary polynomial 
functions. {F}rom (\ref{vleQ}) and (\ref{vleeqn}) we obtain
$P^a_i = -\mfourth\eps_{ijk}\tD^{ja}f^k$ and
\be
X = V_f = f^i\d_i -\fourth(-)^f\eps_{ijk}\tD^i_af^j\tD^{ka},
\ee
where $f = f^i(\th,u)\d_i$ is a vector field acting on $\CC^{3|6}$ and 
$(-)^f = +1$ if $f$ is an even vector field, i.e. $f^i$ is an even 
function.
Due to the symmetry condition (\ref{vlesymm}), the 
components $f^i$ are not independent, but subject to the relation
\be
\tD^{ia}f^j = -\tD^{ja}f^i,
\label{vlefsymm}
\ee

A $\vle(3|6)$ bracket is defined by $[V_f, V_g] = V_{[f,g]_{\vle}}$. 
We find
\bes
[f,g]_{\vle} &=& f^i\d_ig^j\d_j - (-)^{fg}g^j\d_jf^i\d_i \nl
&&+ (-)^f H^m_{ik|jl} \tD^{ia}f^k\tD^j_ag^l\d_m
- (-)^{fg+g} H^m_{jl|ik} \tD^{ja}g^l\tD^i_af^k\d_m, \nl
H^m_{ik|jl} &=& \fourth\eps_{ikj}\dlt^m_l
+\hbox{$1\/16$}\eps_{ijl}\dlt^m_k
+\hbox{$1\/16$}\eps_{jkl}\dlt^m_i.
\ees 
Finally we list the vector fields $f$ and $V_f$ for $\oj_0\ltimes\oj_-$:
\barr{|r|l|l|} 
\hline
\deg&f&V_f\\
\hline
-2& \d_i & E_i \\
-1& -2\eps^{ijk}\th^a_j\d_k & D^{ia} \\
0& u^i\d_j - \third u^k\d_k\dlt^i_j - \eps^{ikl}\th^a_j\th_{ka}\d_l 
& I^i_j\\
0& \eps^{ijk}\th^a_i\th_{jb}\d_k & J^a_b \\
0 & 2u^i\d_i & Z \\
\hline
\earr
\ee
Other vector fields $f$ of non-positive degree are not compatible with the
symmetry condition (\ref{vlefsymm}).

\section{$\mb(3|8)$}
Let $u = (u^i)\in\CC^3$, $i= 1,2,3$, and 
let $z=(z^a)\in\CC^2$, $a=1,2$.
Let $\xi = \xi^i(u)\d/\d u^i$ be a vector field,
let $X = X^a_b(u) T^b_a$ be an $sl(2)$-valued current,
let $\om = \om^i_a(u)\d/\du^i\dz^a$ be a $\CC^2$-valued vector field,
and let $\si = \si^a(u)\d/\d z^a$ be a $\CC^2$-valued function.
Here $T^a_b \in sl(2)$:
$[T^a_b, T^c_d] = \dlt^c_b T^a_d - \dlt^a_d T^c_b$,
and $X^a_a = 0$.
$\mb(3|8)$ has generators $\L_\xi=\L_i(\xi^i)$, $\J_X=\J^b_a(X^a_b)$,
$\G_\om=\G^a_i(\om^i_a)$ and $\S_\si=\S_a(\si^a)$. 
The brackets read \cite{CK99}
\bes
[\L_\xi,\L_\eta] &=& \L_{[\xi,\eta]}
 = \L_k(\xi^i\d_i\yk - \yj\d_j\xi^k), \nl
{[}\L_\xi,\J_X] &=& \J_{\xi X} = \J^b_a(\xi^i\d_i X^a_b),\nl
{[}\J_X,\J_Y] &=& \J_{[X,Y]} = \J^b_a(Y^a_cX^c_b - X^a_cY^c_b),\nl
{[}\L_\xi, \G_\om] &=& 
\G^a_j( \xi^i\d_i\om^j_a + \half\d_i\xi^i\om^j_a - \d_i\xi^j\om^i_a), \nl
{[}\L_\xi, \S_\si] &=& 
\S_a( \xi^i\d_i\si^a + \half\d_i\xi^i\si^a), \\
{[}\J_X, \G_\om] &=&  \G^a_i( X^b_a\om^i_b), \nl
{[}\J_X, \S_\si] &=&  -\S_a( X^a_b\si^b), \nl
\{\G_\om, \G_\ups\} &=&	0, \nl
\{\G_\om, \S_\si\} &=& \L_i( \om^i_a \si^a) 
+ \J^b_a(\d_i\om^i_b\si^a - \om^i_b\d_i\si^a), \nl
\{\S_\si, \S_\tau\} &=& 
 \eps^{ijk}\eps_{ab} \L_k( \d_i\si^a\d_j\tau^b). 
\eens

Consider $\CC^{3|8}$ with basis spanned by three even coordinates
$u^i$, $i=1,2,3$, six odd coordinates $\th_{ia}$, and two
more odd coordinates $\vth^a$.
Let $\deg \th_{ia} = 1$, $\deg u^i = 2$ and $\deg \vth^a = 3$.
The graded Heisenberg algebra has the non-zero relations
\be
\{\eth_b, \vth^a\} = \dlt^a_b, \qquad
[\d_j, u^i] = \dlt^i_j, \qquad 
\{d^{ia}, \th_{jb}\} = \dlt^i_j\dlt^a_b,
\ee
where $\d_i = \d/\d u^i$, $d^{ia} = \d/\d\th_{ia}$, and 
$\eth_a = \d/\d\vth_a$.
The non-positive part of $\mb(3|8)\subset\vect(3|8)$ is spanned by
the vector fields
\barr{|r|l|}
\hline
\deg&\hbox{vector field}\\
\hline
-3 & F_a = \eth_a \\
-2 & E_i = \d_i + \th^a_i\eth_a \\
-1 & D^{ia} = d^{ia} + 3\eps^{ijk}\th^a_j\d_k
+ \eps^{ijk}\th^a_j\th^b_k\eth_b + u^i\eth^a \\
0& I^k_l = u^k \d_l - \th_{la} d^{ka} 
 - \third\dlt^k_l( \ud - \thd ) \\
0& J^c_d = \vth^c \eth_d - \th_{id} d^{ic} 
- \half\dlt^c_d( \veth - \thd ) \\
0& Z = 3 \veth + 2 \ud + \thd\\
\hline
\earr
\label{mbfields}
\ee
The non-zero brackets in $\oj_-$ are
\bes
\{D^{ia}, D^{jb}\} &=& 6\eps^{ijk}\eps^{ab}E_k, 
\nlb{mbcliff}
[D^{ia}, E_j] &=& - 2\dlt^i_j F^a.
\eens
To verify these brackets one must make use of (\ref{skew}), and 
the Jacobi identity $\{\{D^{ia}, D^{jb}\}, D^{kc}\} + \hbox{cyclic} = 0$
requires (\ref{ident}).

A basis for $\toj_{-1}\oplus\toj_{-2}$ is given by
\bes
\tD^{ia} &=& d^{ia} - 3\eps^{ijk}\th^a_j\d_k
+ \eps^{ijk}\th^a_j\th^b_k\eth_b - u^i\eth^a, 
\nlb{mbDE}
\tE_i &=& \d_i - \th^a_i\eth_a, 
\eens
which satisfy
\bes
\{\tD^{ia}, \tD^{jb}\} &=& -6\eps^{ijk}\eps^{ab}\tE_k, \nl
{[}\tD^{ia}, \tE_j] &=& 2\dlt^i_j F^a, 
\label{mbEF}\\
\{D^{ia}, \tD^{jb}\} &=& [D^{ia}, \tE_j] = [\tD^{ia}, E_j] = 0.
\eens
Any vector field in $\vect(3|8)$ has the form 
\be
X = R^a\eth_a + Q^i\d_i + P_{ia}d^{ia}
= \tR^a\eth_a + \tQ^i\tE_i + P_{ia}\tD^{ia},
\label{mbRQP}
\ee
where
\bes
\tQ^i &=& Q^i + 3(-)^X\eps^{ijk}\th^a_jP_{ka}, 
\nlb{mbQR}
\tR^a &=& R^a + (-)^X\th^a_iQ^i 
+ 2\eps^{ijk}\th^a_i\th^b_jP_{kb} - u^iP^a_i.
\eens
$X$ preserves the dual Pfaff equation $\tD^{ia}=0$, i.e.
\be
[X, \tD^{ia}] = -(-)^X \tD^{ia}P_{jb}\tD^{jb},
\ee
provided that
\bes
\tD^{ia}\tQ^j &=& 6(-)^X\eps^{ijk}P^a_k,
\nlb{mbeqn}
\tD^{ia}\tR^b &=& -2(-)^X\eps^{ab}\tQ^i.
\eens
In particular, we have the symmetry relations
\bes
\tD^{ia}\tQ^j &=& -\tD^{ja}\tQ^i,
\label{mbsymm}\\
\tD^{ia}\tR^b &=& -\tD^{ib}\tR^b.
\eens
Compatibility between 
\be
[\tE_i, X] = (\tE_i\tR^a+2P^a_i)F_a + \tE_i\tQ^j \tE_j 
+ \tE_iP_{jb}\tD^{jb}
\label{mbEX}
\ee
and (\ref{mbEF}) in the form 
$\tE_i = {1\/24}\eps_{ijk}\eps_{ab}\{\tD^{ja}, \tD^{kb}\}$, implies
that
\bes
\tE_i\tR^a &=& -2P^a_i, 
\nlb{mb2eqn}
\tE_i\tQ^j &=& \mhalf(\tD^{ka}P_{ka}\dlt^j_i - \tD^{ja}P_{ia}).
\eens
In particular,
\be
\tE_i\tQ^i  = (-)^X \tD^{ia}P_{ia}.
\label{mb2div}
\ee
Hence (\ref{mbEX}) can be written as
\be
[\tE_i, X] = \mhalf(\tD^{ja}P_{ja}\tE_i - \tD^{ja}P_{ia}\tE_j)
+ \tE_iP_{jb}\tD^{jb}.
\ee
However, (\ref{mb2eqn}) are identities which follow from
(\ref{mbeqn}) by considering 
\allowbreak $\eps_{ijl}\eps_{ab}\tD^{ia}\tD^{jb}\tQ^k$ and
\allowbreak $\eps_{ijk}\eps_{ab}\tD^{ia}\tD^{jb}\tR^c$,
so no new independent conditions on the vector fields arise.

Since $\mb(3|8)$ is of depth $3$, there are further relations.
Compatibility between
\be
[F_a,  X] = \eth_a\tR^bF_b + \eth_a\tQ^j \tE_j + \eth_aP_{jb}\tD^{jb}
\ee
and (\ref{mbEF}) in the form $F^a = \sixth[\tD^{ia},\tE_i]$ leads to
\bes
6\eth_a\tQ^i &=& 6(-)^X\eps^{ijk}\tE_jP_{ka} + \tD^j_a\tE_j\tQ^i, 
\nlb{mb3eqn}
3\eth_b\tR^a &=& (-)^X\dlt^a_b\tE_i\tQ^i - \tD^i_bP^a_i.
\eens
In view of (\ref{mb2div}), this means in particular
\be
(-)^X\eth_a\tR^a = \tE_i\tQ^i  = (-)^X \tD^{ia}P_{ia}.
\label{mbdiv}
\ee
Also (\ref{mb3eqn}) are identities which follow from 
(\ref{mbEF}),  (\ref{mbeqn}) and (\ref{mb2eqn}), so no further 
independent conditions on the vector fields arise.

The pairings $\langle \tD^{ia}, \al^j\rangle =
\langle \tD^{ia}, \bt^b\rangle = 0$ give
\bes
\al^i &=& \du^i + 3\eps^{ijk}\th^a_j\dth_{ka}, 
\nlb{mbPfaff}
\bt^a &=& \dvth^a - u^i\dth^a_i + \th^a_i\du^i 
+ 2\eps^{ijk}\th^a_i\th^b_j\dth_{kb}.
\eens
We now show that $\al^i$ and $\bt^a$ satisfy Pfaff equations:
\bes
\L_X \th_{ia} &=& P_{ia}, \nl
\L_X u^i &=& Q^i, \nl
\L_X \vth^a &=& R^a, \nle
\L_X \dth_{ia} &=& (-)^X d^{jb}P_{ia}\dth_{jb} + \d_jP_{ia}\du^j
+ (-)^X\eth_bP_{ia}\dvth^b, \nl
\L_X \du^i &=& -(-)^X d^{jb}Q^i\dth_{jb} + \d_jQ^i\du^j
- (-)^X\eth_bQ^i\dvth^b \nl
\L_X \dvth^a &=& (-)^X d^{jb}R^a\dth_{jb} + \d_jR^a\du^j
+ (-)^X\eth_bR^a\dvth^b.
\eens
In particular,
\barr{lll}
F_c\vth^a = \dlt^a_c, &\ &F_c\dvth^a = 0, \\
F_c\th_{ia} =F_c\dth_{ia} = 0,
&&F_cu^i = F_c\du^i = 0,\\
E_k\vth^a = \th^a_k,
&&E_k\dvth^a = \dth^a_k,\\
E_ku^i = \dlt^i_k, 
&&E_k\du^i = 0, \\
E_k\th_{ia} = 0,
&&E_k\dth_{ia} = 0, \\
D^{kc}\vth^a = \eps^{kij}\th^c_i\th^a_j + \eps^{ca}u^k, 
&&D^{kc}\dvth^a = \eps^{kij}(\th^a_i\dth^c_j + \th^c_i\dth^a_j) 
+ \eps^{ca}du^k, \\ 
D^{kc}u^i = 3\eps^{ikl}\th^c_l, 
&&D^{kc}\du^i = 3\eps^{ikl}\dth^c_l, \\
D^{kc}\th_{ia} = \dlt^k_i\dlt^c_a, 
&&D^{kc}\dth_{ia} = 0, \\
I^k_l\vth^a = 0, 
&&I^k_l\dvth^a = 0, \\ 
I^k_lu^i = \dlt^i_l u^k - \third \dlt^k_l u^i,
&&I^k_l\du^i = \dlt^i_l \du^k - \third \dlt^k_l \du^i, \\
I^k_l\th_{ia} = -\dlt^k_i\th_{la} + \third\dlt^k_l\th_{ia}, 
&&I^k_l\dth_{ia} = -\dlt^k_i\dth_{la} + \third\dlt^k_l\dth_{ia},\\
J^c_d\vth^a = \dlt^a_d\vth^c - \half\dlt^c_d\vth^a, 
&&J^c_d\dvth^a = \dlt^a_d\dvth^c - \half\dlt^c_d\dvth^a, \\
J^c_d u^i = 0, 
&&J^c_d \du^i = 0, \\
J^c_d\th_{ia} = -\dlt^c_a\th_{id} + \half\dlt^c_d\th_{ia}, 
&&J^c_d\dth_{ia} = -\dlt^c_a\dth_{id} + \half\dlt^c_d\dth_{ia}, \\
Z\vth^a = 3\vth^a, 
&& Z\dvth^a = 3\dvth^a, \\
Zu^i = 2u^i, 
&&Z\du^i = 2\du^i, \\
Z\th_{ia} = \th_{ia}, 
&& Z\dth_{ia} = \dth_{ia}.
\earr
\ee
One checks that
\barr{lll}
I^k_l\al^i = \dlt^i_l \al^k - \third \dlt^k_l \al^i,
&&I^k_l\bt^a = 0, \\
J^c_d\al^i = 0,
&&J^c_d\bt^a = \dlt^a_d\bt^c - \half\dlt^c_d\bt^a,  \\
Z \al^i = 2\al^i,
&&Z\bt^a = 3\bt^a, \\
D^{kc}\al^i = E_k\al^i = F_c\al^i = 0,
&&D^{kc}\bt^a = E_k\bt^a = F_c\bt^a = 0.
\earr
\ee
Thus the systems of Pfaff equations $\bt^a=0$ and
$\al^i = \bt^a = 0$ are preserved by 
$\oj_-$ and $\oj_0$ and therefore by all of $\mb(3|8)$. 
A long calculation gives
\bes
\L_X\al^i &=& \tE_j\tQ^i\al^j - (-)^X\eth_a\tQ^i\bt^a, 
\nlb{mbXab}
\L_X\bt^a &=& (-)^X\eth_b\tR^a\bt^b,
\eens
provided that the conditions (\ref{mbeqn}) and (\ref{mb2eqn}) hold.

Hence $\mb(3|8)$ has three natural classes of tensor modules, 
with bases $\al^i$, $\bt^a$ and $\gm^{ia}$, and module action
\bes
\L_X\al^i &=& \tE_j\tQ^i\al^j - (-)^X\eth_a\tQ^i\bt^a, \nl
\L_X\bt^a &=& (-)^X\eth_b\tR^a\bt^b, 
\label{mbmod}\\
\L_X\gm^{ia} &=& -(-)^X \tD^{ia}P_{jb}\gm^{jb}.
\eens
One notes that the homogeneous part of the first equation must define
a module action by itself, since $\bt^a$ transforms homogeneously. Hence
we set
\be
\L_X\al^i &=& \tE_j\tQ^i\al^j.
\ee
Let us assume that $\al^i$ and $\bt^a$ are fermionic. The volume forms
$v_\al = \al^1\al^2\al^3$, $v_\bt = \bt^1\bt^2$ and 
$v_\gm = \gm^{11}\gm^{12}\gm^{21}\gm^{22}\gm^{31}\gm^{32}$ transform as
\bes
\L_Xv_\al &=& \tE_i\tQ^i v_\al, \nl
\L_Xv_\bt &=& (-)^X\eth_a\tR^a v_\bt,\\
\L_Xv_\gm &=& -(-)^X \tD^{ia}P_{ia} v_\gm.
\eens
Since $v_\al$ and $v_\bt$ have degree $+6$ each, and $v_\gm$ has degree
$-6$, the forms $v_\al v_\gm$ and $v_\bt v_\gm$ are invariant, which 
implies the relations (\ref{mbdiv}).

{F}rom (\ref{mbmod}) we deduce the transformation laws for a scalar 
density $v$ of weight $+1$, an $sl(3)$ vector $\bal^i$, an $sl(2)$ 
vector $\bbt^a$, and an $sl(3)\ooplus sl(2)$ vector $\bgm^{ia}$:
\bes
\L_Xv &=& \sixth\d_i\tQ^i v 
= \msixth\tD^{ia}P_{ia}v, \nl
\L_X\bal^i &=& 
\msixth(\dlt^i_j\tD^{ka}P_{ka} - 3\tD^{ia}P_{ja})\bal^j\nl
&=& (\tE_j\tQ^i - \third\dlt^i_j\tE_k\tQ^k)\bal^j, \nle
\L_X\bbt^a &=& \msixth(\dlt^a_b\tD^{ic}P_{ic} - 2\tD^{ia}P_{ib})\bbt^b \nl
&=& (-)^X(\eth_b\tR^a - \half\dlt^a_b\eth_c\tR^c)\bbt^b, \nl
\L_X\bgm^{ia} &=& \msixth(\dlt^i_j\dlt^a_b\tD^{kc}P_{kc}
-6\tD^{ia}P_{jb})\bgm^{jb}.
\eens
Hence we obtain the explicit realization
\bes
\L_X &=& X + \msixth \tD^{ia}P_{jb}(-3\dlt^b_a\bI^j_i - 2\dlt^j_i\bJ^b_a
+ \dlt^j_i\dlt^b_a\bZ) 
\nlb{mbtensor}
&=& X + \tE_j\tQ^i\bI^j_i + (-)^X\eth_b\tR^a\bJ^b_a + \sixth\tE_i\tQ^i \bZ,
\eens
where $\bI^i_j$, $\bJ^a_b$ and $\bZ$ generate the Lie algebra 
$\boj_0 = \ssg$ and $X$ commutes with $\boj_0$. 
By introducing canonical conjugate oscillators $\bal^*_i$, $\bbt^*_a$
and $v^*$, subject to the Heisenberg algebra
$[\bal^*_i, \bal^j] = \dlt^j_i$, $[\bbt^*_a, \bbt^b] = \dlt^b_a$, 
$[v^*, v] = 1$,
we obtain the explicit expression for the $\boj_0$ generators: 
$\bI^i_j = \bal^i\bal^*_j - \third\dlt^i_j\bal^k\bal^*_k$, 
$\bJ^a_b = \bbt^a\bbt^*_b - \half\dlt^a_b\bbt^c\bbt^*_c$ and
$\bZ = vv^*$. However, it is clear that $\L_X$ 
will satisfy the same algebra for every representation of $\ssg$.
Substitution of irreducible $\boj_0$ modules into (\ref{mbtensor}) 
gives the tensor modules for $\mb(3|8)$. In particular, the expressions
for $\oj_0$ become $\L_{I^i_j} = I^i_j + \bI^i_j$,
$\L_{J^a_b} = J^a_b + \bJ^a_b$ and $\L_Z = Z + \bZ$,
i.e. two commuting copies of $\oj_0$.

To obtain an explicit expression for the vector fields in $\mb(3|8)$,
we set $\tR^a = f^a(\th,u,\vth)$, two arbitrary polynomial 
functions. {F}rom (\ref{mbQR}) and (\ref{mbeqn}) we obtain
\bes
\tQ^i &=& \mfourth\tD^{ia}f_a, \nle
P^a_i &=& \hbox{${1\/48}$}\eps_{ijk}\tD^{ja}\tD^{kb}f_b, 
\eens
i.e.
\be
X &\equiv& M_f = f^a\eth_a +\fourth(-)^f\tD^{ia}f_a\tE_i
+ \hbox{${1\/48}$}\eps_{ijk}\tD^i_a\tD^{jb}f_b\tD^{ka},
\label{mbX}
\ee
where $f = f^a(\th,u,\vth)\eth_a$ is a vector field acting on 
$\CC^{3|8}$ and $(-)^f = +1$ if $f$ is an even vector field, 
i.e. $f^a$ is an odd function.
Due to the symmetry condition (\ref{mbsymm}),
the components $f^a$ are not independent, but subject to the relations
\bes
\tD^{ia}f^b &=& -\tD^{ib}f^a, 
\nlb{mbfsymm}
\tD^{ia}\tD^{jb}f_b &=& - \tD^{ja}\tD^{ib}f_b.
\eens

$\mb(3|8)$ contains an $\vle(3|6)$ subalgebra. We see from 
(\ref{mbXab}) that if $\eth_b\tQ^a = 0$, then the Pfaff equation
$\al^i=0$ is preserved, not just the combined Pfaff equations
$\al^i=\bt^a = 0$. This condition defines a subalgebra since 
additional structure is preserved. $\tQ^i$ and $P_{ia}$ can now
be expressed in terms of three $\vth$-independent functions 
\be
f^i(\th,u)= \mfourth \tD^{ia}f_a(\th,u,\vth),
\label{fi}
\ee
as $\tQ^i = f^i$ and $P^a_i = {1\/48}\eps_{ijk}\tD^{ja}f^k$.
It is clear from (\ref{mbRQP}) that when $\tQ^i$ and $P_{ia}$ 
do not depend on $\vth$, the bracket between two vector
fields is completely determined by the part 
$X' = \tQ^i\d_i + P_{ia}\tD^{ia}$, and hence the subalgebra is 
$\vle(3|6)$. Note that $f^a$ still depends on $\vth$, in the way specified
by (\ref{fi}).

An $\mb(3|8)$ bracket is defined by $[M_f, M_g] = M_{[f,g]_{\mb}}$. 
We find
\bes
[f,g]_{\mb} &=& f^a\d_ag^b\eth_b - (-)^{fg}g^b\d_bf^a\eth_a \nl
&&+ \fourth(-)^f\tD^{ia}f_a\tE_ig^b\eth_b
- \fourth(-)^{fg+g}\tD^{jb}g_b\tE_jf^a\eth_a \nl
&&+ \eps_{ikj}H^e_{abd} \tD^{ia}\tD^{kc}f_c\tD^{jb}g^d\eth_e \\
&&-(-)^{fg}\eps_{jli}H^e_{bad}
 \tD^{jb}\tD^{lc}g_c\tD^{ia}f^d\eth_e, \nl
H^e_{abd} &=& -\hbox{${1\/48}$}\eps_{ab}\dlt^d_e 
- \hbox{${1\/96}$}\eps_{bd}\dlt^e_a.
\eens 

Finally we list some of the vector fields $f$ and $M_f$ for 
$\oj_0\ltimes\oj_-$:
\barr{|r|l|l|} 
\hline
\deg&f&M_f\\
\hline
-3& \eth_a & F_a \\
-2& 2\th^a_i\eth_a & E_i \\
-1& 2u^i\eth^a + 6\eps^{ijk}\th^a_j\th^b_k\eth_b & D^{ia} \\
0 & 3\vth^a\eth_a + u^i\th^a_i\eth_a & Z \\
\hline
\earr
\ee
To compute the vector fields $f$ which correspond to $I^i_j$ and 
$J^a_b$ is quite tedious and has not been attempted. However, 
since $\eps^{ijk}\th^a_j\th_{ka}\equiv 0$ it is
clear that $M_f = I^i_j$ for $f$ some linear combination of
$u^i\th^a_j\eth_a$ and $\eps^{ikl}\th^a_j\th_{ka}\th^b_l\eth_b$,
and $M_f = J^a_b$ for $f$ some linear combination of
$\vth^a\eth_b$, $u^i\th^a_i\eth_b$ and
$\eps^{ijk}\th^a_i\th_{bj}\th^c_k\eth_c$; 
the right combinations are found
by demanding that the symmetry conditions (\ref{mbfsymm}) hold.
Other vector fields of non-positive degree are not compatible with
(\ref{mbfsymm}).

\section{Acknowledgement}
I am grateful to D. Leites for discussions and for giving me preprints
before publication. I first heard about the exceptional Lie superalgebras
and their possible connection to the standard model in talks by 
V. Kac and A. Rudakov at the Fields Institute in the fall of 2000. I 
thank S. Berman and Y. Billig for inviting me there.

\end{document}